# Sparse Convolution-based Markov Models for Nonlinear Fluid Flows

Chen Lu[1], Balaji Jayaraman[2,*]

*School of Mechanical and Aerospace Engineering, Oklahoma State University, Stillwater, Oklahoma, USA*

Joshua Whitman[1], Girish Chowdhary[2]

*Aerospace Engineering, University of Illinois, Urbana-Champaign, Illinois, USA*

**Abstract**

Data-driven modeling for nonlinear fluid flows using sparse convolution-based mapping into a feature space where the dynamics are Markov linear is explored in this article. The underlying principle of low-order models for fluid systems is identifying convolutions to a feature space where the system evolution (a) is simpler and efficient to model and (b) the predictions can be reconstructed accurately through deconvolution. Such methods are useful when real-time models from sensor data are needed for online decision making. The Markov linear approximation is popular as it allows us to leverage the vast linear systems machinery. Examples include the Koopman operator approximation techniques and evolutionary kernel methods in machine learning. The success of these models in approximating nonlinear dynamical systems is tied to the effectiveness of the convolution map in accomplishing both (a) and (b) mentioned above. To assess this, we perform in-depth study of two classes of sparse convolution operators: (i) a pure data-driven POD-convolution that uses left singular vectors of the data snapshots - a staple of Koopman approximation methods and (ii) a sparse Gaussian Process (sGP) convolution that combines sparse sampling with a Gaussian kernel embedding an implicit feature map to an inner product reproducing kernel Hilbert space. We are particularly interested in the effectiveness of these convolution maps for long-term prediction using limited data for three classes of fluid flows with escalating complexity starting from a limit-cycle attractor in a cylinder wake followed by a transient evolution into an attractor and finally, a highly transient buoyant Boussinesq mixing flow. The results indicate that a purely data-driven POD-convolution framework optimally maps information from input to the feature space, but is limited for highly transient flows when the singular vectors get outdated. On the other hand, the more generic sparse Gaussian Process (sGP)-based convolution maps are less sensitive to the evolution of the data but introduce deconvolution errors. Both methods show mixed outcomes for the effectiveness of the Markov linear model in feature space. We also explore layering of convolution operators as a way to bypass the high dimensionality of nonlinear dictionary-based convolution maps which is shown to be particularly effective for the sGP framework.

---

[*]Corresponding author
  *Email address:* `balaji.jayaraman@okstate.edu` (Balaji Jayaraman)
[1]Graduate Student
[2]Assistant Professor

# 1. Introduction

Nonlinear fluid flows are highly complex, spatiotemporally evolving multi-scale dynamical systems that are difficult to model even with knowledge of the governing equations. The challenge in modeling such systems is the multiscale nature that renders them high-dimensional. Reducing the dimensionality is challenging especially when the smaller-scale structures with low energy content may still turn out to be dynamically important. This challenge is magnified when the governing equations are unknown. However, there still exists a strong demand for purely data-driven models for control, online analysis and decision making from fluid flow data streaming in from sensors or particle image velocimetry (PIV). This paper contributes to the literature on sparse convolution-based models for analyzing and predicting nonlinear fluid flow dynamical systems accurately and efficiently using a limited amount of data. Sparse representation is inherently associated with dimensionality reduction so that models can be developed efficiently in real-time using onboard computing power. The convolution part of the model serves two purposes: (i) provides a sparse low-dimensional basis that spans the data; (ii) maps to a feature space where the evolution is Markovian and hopefully, linear.

Proper orthogonal decomposition (POD) [1] is popular for dimensionality reduction which when combined with Galerkin projection [2, 3] onto the flow partial differential equations (PDEs) allows for the prediction of the system in a low-dimensional basis space through simpler ordinary differential equations (ODEs). The underlying principle being that the evolution of the weights in a feature space can approximate the system behavior in the physical space as long as the mapping remains valid. This is equally applicable to both data and system of equations. Extracting high-fidelity models of a dynamical system from snapshots in time is a major need in engineering where sensor data is widely available. It may be advantageous to learn the model from evolution in a low-dimensional feature space, commonly modeled as a Markov linear process which allows us to leverage the vast literature on linear systems control [4], optimization and spectral analysis. Such approaches primarily evolved as system identification tools [5] but has recently become popular for spectral analysis of fluid flows using data [6]. For a given current state $\boldsymbol{x}_t$, and future state $\boldsymbol{x}_{t+T}$, a Markov model [7] is given by $\boldsymbol{g}(\boldsymbol{x}_{t+T}) = \mathcal{K}\boldsymbol{h}(\boldsymbol{x}_t)$ where $\boldsymbol{g}$ and $\boldsymbol{h}$ are typically finite-dimensional transformations to the feature space and $\mathcal{K}$ represents the linear transition operator. As will be shown section 2, such a Markov process can approximate the Koopman operator [8, 9, 6] under certain conditions. In the Koopman framework, the feature space is the observable space and the feature maps are observable functions.

Dynamic mode decomposition(DMD) [10, 6] is a linear model in the input space by employing feature transformations that are linear, identical map of the state vectors themselves, i.e. $\boldsymbol{g}(\boldsymbol{y}) = \boldsymbol{y}$ and $\boldsymbol{h}(\boldsymbol{x}) = \boldsymbol{x}$. The DMD approximates the Koopman operator [9] to the extent that it captures a subset of the Koopman modes [6, 11, 12]. It represents a decomposition obtained through projection onto eigenfunctions of the linear Koopman operator which represents a Markovian linear evolution of the underlying system in the space of observables. Thus one could obtain valuable physical insights from knowledge of dynamically relevant modes oscillating at a frequency and grow or decay in amplitude corresponding to the eigenvalues of the Koopman operator [6].

In machine learning, it is common to mine for similarity measures [13, 14] within the underlying data that can then be leveraged for low-dimensional representation by embedding generic positive definite kernels at optimally chosen locations. These kernels represent dot products of infinite-dimensional feature (convolution) maps in an inner-product space. Operating in this dot product space allows one to build efficient learning algorithms through the *kernel trick*. Such a positive definite kernel implicitly defines the convolution map from Mercer's theorem. Thus, the spattiotemporal data can be represented by the kernel coefficients and their dynamics. An example of such an approach is the Evolving Gaussian Process or EGP [15, 16] where one models the spatiotemporal process as an evolution of the Gaussian kernel weights in a function space, typically, but not necessarily as Markovian transitions. The Gaussian kernel here implicitly represents a feature map, while dimensionality reduction is achieved using a sparse measurement of the kernel matrix.

There exist a whole host of similar Markov modeling approaches including the Variational Approach for Markov Processes (VAMP) [7], Markov State Models [17] and Ulam's Galerkin Method [18]. All these variants build a finite-dimensional Markov linear model in a feature space. Often, it is *not always clear what the optimal transformations to the feature space are*. The variational approach-based methods such as VAMP and Variational Approach to Conformation Dynamics (VAC) [19] learn optimal feature map from amongst those spanned by a dictionary of basis functions (e.g. monomials, radial basis functions or family of orthogonal polynomials). The Extended DMD or EDMD [20] can be viewed as a specific case of this class of methods where the basis functions are polynomials. The downside is that the model will still depend on the basis dictionary which can rapidly grow in dimension with complexity. To reduce dimensionality, EDMD was adapted to include the kernel trick to build Kernel DMD or KDMD [21]. The *Sparse Identification of Nonlinear Dynamical Systems* or SINDy [22] is another variant that employs $L_1$ regression to build the sparsest basis representation to approximate the Koopman eigenfunctions.

All of the above techniques combine elements of sparse representation with a transformation map, $g$, $h$ to feature space where the dynamics are easier to model. Setting $g = h$ makes this Markov linear model into a finite-dimensional approximation of the Koopman operator[8, 23, 6] which is a composition operator. By this, one implies that the Koopman operator evolves the observables in the feature space through its composition with the evolution operator in input space. All of the above discussion essentially boils the data-driven modeling problem down to identifying the appropriate feature map to a linear system space where the evolution operator (Koopman) can easily be learned from the wide array of regression techniques. There are three broad classes of feature maps: (C1) purely data-driven map from training data; (C2) an implicit map to a higher dimensional dot product feature space through an appropriate positive definite kernel; and (C3) basis function dictionary from which the optimal choice for the given training data can be identified. All these approaches incorporate dimensionality reduction either in the feature space or in the input space or both. An example for (C1) is a convolution map that projects the state onto left singular vectors of the training data and operates on the features. An example for (C2) is to combine sparse measurement in the input space with a feature map embedded in a Gaussian Kernel [24, 25, 26] as a sparse Gaussian Process (sGP) convolution. This is attractive for building models from a limited sensor data as is typical in experimental studies. In such methods, the structure of the measurement matrix (sensor placement) impact the prediction quality. Examples for (C3) include VAMP, EDMD, and SINDy, but is hardly ever applied directly to high-dimensional fluid flow data, but in conjunction with dimensionality reduction techniques.

In this paper, we explore the impact of different classes of feature maps on data-driven Markov modeling methods so as to provide a unified perspective of their capabilities for modeling fluid flows. A secondary motivation for this study is to ultimately enable cross-pollination of ideas to build improved models. The success of the sparse convolution Markov models ultimately hinge on their ability to accomplish two things: (a) to map across the input and feature spaces accurately and efficiently and (b) to predict the dynamical evolution of the system in the feature space accurately and efficiently. With this mind, we assess and compare the two very different classes of data-driven sparse convolution operators: (i) a data-driven POD-convolution approach similar to (C1) using truncated singular value decomposition (SVD) that is a staple of Koopman approximation methods and (ii) a sparse Gaussian Process (sGP) convolution (C2) that combines a sparse measurement matrix with a feature map embedded in a Gaussian kernel. The assessment emphasizes long-term prediction metrics from limited training data of canonical fluid flows with varying levels of complexity including (i) a limit-cycle attractor in a cylinder wake, (ii) wake instability involving transient evolution into an attractor and (iii) highly transient buoyancy-driven mixing flow. To embed nonlinearity in the convolution maps and reduce the dimension of the system, we explore the idea of layering multiple convolution operators layered sequentially and optimized locally. We term this approach as the *multilayer convolution*. In this study, we limit these architectures to two layers although efforts to extend this to arbitrary numbers of layers and optimizing them non-locally are currently being pursued. For comparison, we also assess these methods with the well-known EDMD [20] and its kernel variant, KDMD [21].

The paper is organized as follows. In section 2, we connect the sparse convolution-based Markov linear modeling to the class of Koopman approximation methods. In section 3.1, we show sparse-convolution as a mapping to the Koopman observable space. In sections 3.2 and 3.3, we show how the different convolution maps (POD- and sGP-convolution) can generate different data-driven predictive models such as the DMD and EGP and extend these to a multilayer convolution framework in section 3.4 as a way to build complex feature maps without paying the cost of high-dimensional features. In section 3.5, we explore different strategies for sensor placement that impacts the structure of the sGP-convolution. In section 4, we present numerical examples to analyze the performance of the various single and multilayer convolution models on canonical fluid flow problems. The major conclusions from this study are summarized in section 5.

## 2. Koopman as a Linear Markov Process

Consider a discrete-time dynamical fluid system with its evolution at any given instant represented as below:
$$\boldsymbol{y} = \boldsymbol{F}(\boldsymbol{x}), \tag{1}$$
where $\boldsymbol{x}, \boldsymbol{y} \in \mathcal{M}$ are $N$-dimensional state vectors, e.g., velocity components at discrete locations in a flow field at a current instant $t$, and separated by an appropriate unit of time $T$. To be precise, $\boldsymbol{x} \triangleq \boldsymbol{x}_t$ and $\boldsymbol{y} \triangleq \boldsymbol{x}_{t+T}$. Operator $\boldsymbol{F}$ describes a non-linear system which evolves state $\boldsymbol{x}$ to state $\boldsymbol{y}$ in time, i.e. $\boldsymbol{F}: \mathcal{M} \to \mathcal{M}$. Note that we can equivalently represent this system in a continuous-time setting. A generalized version of a linear Markov model [7] for such a dynamical system is shown below:
$$\boldsymbol{g}(\boldsymbol{y}) = \boldsymbol{g}(\boldsymbol{x}_{t+T}) = \mathcal{K}\boldsymbol{h}(\boldsymbol{x}_t) = \mathcal{K}\boldsymbol{h}(\boldsymbol{x}). \tag{2}$$
Here, $\boldsymbol{g}(\boldsymbol{y}) = (g_1(\boldsymbol{y}), g_2(\boldsymbol{y}), g_3(\boldsymbol{y}), ...)^T$ and $\boldsymbol{h}(\boldsymbol{y}) = (h_1(\boldsymbol{y}), h_2(\boldsymbol{y}), h_3(\boldsymbol{y}), ...)^T$ are vector-valued, i.e. column vectors with scalar-valued basis functions as its elements. These elements represent the components of the transformation of the state vector into a feature space, $\mathbb{C}$ or $\boldsymbol{g}(\boldsymbol{y}), \boldsymbol{h}(\boldsymbol{x}) \in \mathbb{C}$. Although, $\boldsymbol{g}, \boldsymbol{h} \in \mathcal{F}$ (where $\mathcal{F}$ is a function space) are usually finite-dimensional approximations in practice, they are infinite-dimensional in theory. Consequently, the linear transition operator $\mathcal{K}$ is always a finite-dimensional approximation of the infinite-dimensional truth. The Koopman operator theoretic view interprets $\mathcal{K}$ as operating on the observable space [20] $\mathcal{K}: \mathcal{F} \to \mathcal{F}$. If the transformations to the feature space $g$ and $h$ are identical and chosen such that the dynamics of the system are linear in the feature space, then $\mathcal{K}$ (Eq. (2)) represents a finite-dimensional approximation of the Koopman operator [8, 9]. DMD [10] is a Koopman approximation method employing a linear identity map between the observable and input spaces and hence, represents a linear model in the input space. The resulting evolutionary model is given by:

$$\mathcal{K}g(x) = g(y) = g(F(x)). \tag{3}$$

Since the Koopman operator has the effect of operating on the functions of state space as shown in Eq. (4), it is commonly referred to as a composition operator with ∘ representing the composition between $g$ and $F$.

$$\mathcal{K}g = g \circ F. \tag{4}$$

Being a linear operator, spectral analysis of $\mathcal{K}$ include the Koopman eigenfunctions ($\phi_j$), eigenmodes ($v_j$) and eigenvalues ($\mu_j$) which can be leveraged to reconstruct the transformation $g(x)$ as shown in Eqs. (5)-(6). Thus, the transformation $g$ should be spanned by Koopman eigenfunctions $\phi$.

$$g(x) = \sum_{j=1}^{\infty} \phi_j v_j, \tag{5}$$

$$g(y) = \mathcal{K}g(x) = \sum_{j=1}^{\infty} \phi_j v_j \mu_j. \tag{6}$$

There exist many methods to approximate the Koopman tuples including DMD [6, 10], EDMD [20], kernel DMD [21] and generalized Laplace analysis (GLA) [23]. In this paper, we show that the various sGP-based sparse convolution models can also accomplish the same. The key to accurately and effectively capturing the dynamical behavior of the nonlinear system relies heavily on the choice of the observable function $g$. The centrality of this paper is to assess effectiveness of sparse convolution acts as feature maps for long-time prediction.

## 3. Sparse Convolution Approximates Koopman Observables

### 3.1. Model Formulation

We define a convolution operation as a projection of the input state onto a basis space such that the dynamics evolve through the feature variables. This would require the state $x$ be spanned accurately by the choice of basis space $g_i$. It is worth noting that such a convolution allows for both functional as well data-driven forms, e.g., POD basis from data or Gaussian basis. From the discussion in section 2, we note that the transformation $g$ should be spanned by Koopman eigenfunctions $\phi$ as in eq. (5) to accurately capture the Koopman tuples. Obviously, the challenge is that $\phi$ is not known *a priori*. Given pairs of snapshot data $X = [x_1, x_2, ..., x_M]$ and $Y = [y_1, y_2, ..., y_M]$, where $X, Y \in \mathbb{R}^{N \times M}$. $N$ is the dimension of the state, and $M$ is the number of data snapshots. We approximate the true non-linear operator $F$, which governs the time evolution from $x_i$ to $y_i$ ($i = 1, 2, 3, ..., M$), using a quasi-linear operator $A(X)$ that maps the snapshot data $X$ to $Y$ in time as below:

$$A(X)X = Y. \tag{7}$$

Using an appropriate observable functions $\boldsymbol{g}$, the system evolves through the Koopman operator as:

$$\mathcal{K}\boldsymbol{g}(X) = \boldsymbol{g}(Y), \tag{8}$$

where $\mathcal{K}$ is linear with dimension tied to that of $\boldsymbol{g}$. To build a finite-dimensional $\mathcal{K}$ in Eq. (8) we define a finite-dimensional convolution operator $C \in \mathbb{R}^{N \times K}$ as a quasi-linear approximation to $\boldsymbol{g}$ such that:

$$X = C(X)\bar{X}, \tag{9}$$

$$Y = C(Y)\bar{Y}, \tag{10}$$

where $\bar{X} \in \mathbb{R}^{K \times M}$ and $\bar{Y} \in \mathbb{R}^{K \times M}$ are the corresponding weights for $X$ and $Y$ in the feature space, and $K$ represents the feature dimension. Note that $C$ can be nonlinear and should evolve with $X$ as $C(X)$. Then, Eq. (7) can be rewritten as:

$$A(X)C(X)\bar{X} = C(Y)\bar{Y}. \tag{11}$$

Rearranging Eq. (11) gives the relationship below with $C^+$ being the Moore-Penrose pseudo-inverse

$$C(Y)^+ A(X) C(X) \bar{X} = \bar{Y}. \tag{12}$$

Defining $\bar{A} \triangleq C(Y)^+ A(X) C(X)$ as the linear operator in the feature space we get:

$$\bar{A}\bar{X} = \bar{Y}. \tag{13}$$

If the approximated model for the observable functions $\boldsymbol{g}(X)$ and $\boldsymbol{g}(Y)$ are appropriate for the dynamics of interest, then $\bar{A}$ will be truly linear. Examining Eqs. (12) and (13), we can see that $\bar{A}$ can be independent of $\bar{X}$ only when $C$ evolves with the state as $C(X)$ and $C^+(X)$ exists. However, in practice $C$ is either predetermined or depends on only on training data which makes $\bar{A}$ as:

$$\bar{A} = C(Y_0)^+ A(X) C(X_0), \tag{14}$$

where $C(X_0), C(Y_0)$ are the based on the training data $X_0, Y_0$. Note that finite-rank approximation of $\mathcal{K}$ is embedded in the dimensions of convolution operator $C \in \mathbb{R}^{N \times K}$, with $K$ considerably smaller than $M$ and $N$. Given $C$, $X$ and $Y$, we can learn $\bar{A}$ by minimizing the Frobenius norm $\|\bar{A}\bar{X} = \bar{Y}\|_F$. In practice, one often uses Tikhonov regularization to seek a unique solution.

### 3.2. POD Convolution as DMD

The DMD algorithm [10] employs a truncated set of singular vectors of the data $X$ to reduce its dimensionality and also of the linear system model Eq. (7) and build the convolution operator $C$. Acknowledging that the singular value decomposition (SVD) and POD differ in whether the mean of the data is removed or not, we refer to this method as POD-convolution. The SVD of $X$ is

$$X = U\Sigma W^T. \tag{15}$$

DMD employs a similarity transform of $A$ using the truncated set of left singular vectors $U$. In the convolution framework, we project $X, Y$ onto $C^+ = U^T$ as in Eq. (9) and (10) and rewrite Eq. (7) as:

$$AU\bar{X} = U\bar{Y}. \tag{16}$$

Using $\bar{Y} = U^T Y$ and $\bar{X} = U^T X = \Sigma W^T$ from Eq. (15) and rearranging, we have:

$$U^T A U = U^T Y W \Sigma^{-1}. \tag{17}$$

Defining the reduced transition operator as $\tilde{A} \triangleq U^T A U$, we have:

$$\tilde{A} = \bar{Y} \bar{X}^+ = U^T Y W \Sigma^{-1}. \tag{18}$$

Similar to Eq. (13), the resulting model of the system is shown in Eq. (19):

$$\tilde{A}\bar{X} = \tilde{A}\Sigma W^T = U^T Y = \bar{Y}, \tag{19}$$

where $\Sigma W^T$ and $U^T Y$ represent $\bar{X}$ and $\bar{Y}$. One can easily see that Eqs. (17)-(19) are the DMD algorithm [10]. The benefit of using POD modes is to generate an optimal low-dimensional basis (dimension $K$) for the given data set (relative to the dimension of the state) while also allowing one to perform exact deconvolution of $C$ through a simple transpose operation, i.e., $C^+ = C^T = U^T$. One can show that the tuples (eigenvalues, eigenfunctions, and eigenmodes) of $\tilde{A}$ learned through the POD-convolution approximates a subset of the Koopman operator. We refer the reader to Rowley et al. [6], Tu et al. [27] and Rowley and Dawson [11] for discussion on the connections between DMD and Koopman theory. Since POD modes are purely data-driven, the linear operator $\bar{A}$ in eq. (19) will approximate the Koopman operator well in regions where data is available, but will not adequately span the flow states in a highly transient regime outside the training region.

### 3.3. Sparse Gaussian Processes Convolution

It is common in the field of machine learning to identify the underlying structure of data so that its representation can be generalized to unknown data points. This requires learning similarity measures as the underlying basis conctituting the convolution map. The SVD basis in section 3.2 is compactly optimal for the given data but not generalizable. Generalization requires functional basis that are hard to design. Commonly, one defines similarity measures between $x$ and $y$ in a feature space [14] through kernel functions, $k(x, y) = \langle \phi(x), \phi(y) \rangle$, where $\phi$ is a map to an inner product space. This approach allows one to build compact algorithms for regression and model learning in this dot product space. There exist many such examples of kernels including Gaussian Processes (GP) [28, 15, 16], Random Kitchen Sinks (RKS) [29], and Fastfood [30]. These kernels serve the same purpose as the POD convolution, namely: (i) provide a compact representation of spatial information; (ii) describe the underlying structure of data; and (iii) convert a spatiotemporal process into a temporal evolution in the feature space. A major difference with such kernel basis is that they are not necessarily optimally compact for the given data, almost always non-orthogonal and hyperparameter sensitive. In this paper, we explore the use of GP kernel functions (Eq. (20)) for building the convolution operators and assess their suitability for modeling fluid flow dynamics.

$$C_{ij} = C(z_i, \bar{z}_j) \triangleq \exp \frac{-||z_i - \bar{z}_j||^2}{2\sigma^2}, \tag{20}$$

In the above we define the elements of the sGP convolution operator, $C \in \mathbb{R}^{N \times K}$ with $i = 1, 2, ..., N$ and $j = 1, 2, ..., K$, $z$ is a vector of spatial locations where all the state information of the fluid flow is available or sensor locations, i.e., $X = X(z, t)$, and $\bar{z}$ is a vector of sparse sensor locations optimally chosen to represent the given dataset. The kernel $C_{ij}$ in Eq. (20) represents a generic spatial similarity measure with $\sigma$ being the model hyperparameter denoting the spreading width of the functions. The centers $\bar{z}_j$ can be identified from analyzing the available snapshot data using one of the many well known techniques including POD-based sensor placement [31], k-means clustering [28] and Sparse Online Gaussian Processes (SOGP) [32] to name a few. For known dynamics, one can place centers at the relevant locations, e.g., extrema of the most energetic POD modes [31]. In this work, we have pursued three different algorithms to learn centers as discussed in section 3.5.

Typically, the number of centers (features) $K$ is much smaller than $N$ in order to build a low-dimensional model. Because the kernel functions themselves are not designed to be orthogonal, inverting the convolution operator $C$ is inexact, requiring the computation of a generalized Moore-Penrose pseudo-inverse for deconvolution. The accuracy of this matrix inversion is sensitive to the sensor locations in the measurement matrix which should be designed to accurately span the input data. In general, the sensor plcament should be such that $C$ has a full column rank. Using sGP convolution is akin to building an evolutionary Markov model from limited sensor data of a complex flow field and represents the core of the Evolving Gaussian Process or EGP class of methods [15, 16].

### 3.4. Multilayer Convolution for Nonlinear Dynamics

The two classes of convolution operators described above can be viewed as data-dependent to varying degrees and do not evolve with the instantaneous state of the system. The basis vectors onto which the convolution projects the input state need to be of a functional form of the instantaneous state. As shown in Eq. (12) and Eq. (13), the convolution operator $C$ should be $C(X)$ so that $\bar{A}$ has a chance to be linear. EDMD [20] and VAMP [7] use a dictionary of functions to build the basis that constitute the convolution operator. However, such approaches are still limited by the composition of the dictionary, susceptible to uncontrolled growth in the feature dimension and not generalizable. In this section, we explore building complex convolution maps through layering of elementary convolution operators. The hypothesis is that employing a *deep* convolution made up of low-dimensional units through layering can offer better feature maps than a higher dimensional single-layer convolution. In other words, a shorter and deeper convolution is better than a taller and shallow map. A generalized way of building convolution operator is to layer recursively multiple convolution operators such as:

$$X = C_1 C_2 C_3 ... C_L \bar{X}_L = \boldsymbol{C}_{ML} \bar{X}_L, \tag{21}$$

$$Y = C_1 C_2 C_3 ... C_L \bar{Y}_L = \boldsymbol{C}_{ML} \bar{Y}_L, \tag{22}$$

where $\bar{X}_L$ and $\bar{Y}_L$ represent the mapped features at the $L^{th}$ layer and $\boldsymbol{C}_{ML}$ represents the multilayer convolution operator obtained by layering individual convolutions such as POD, sGP and other transfer functions. Substituting Eq. (21) and (22) into Eq. (13), we have:

$$A\boldsymbol{C}_{ML}\bar{X} = \boldsymbol{C}_{ML}\bar{Y}. \tag{23}$$

Pre-multiplying the pseudoinverse of $\boldsymbol{C}_{ML}$, we have:

$$\boldsymbol{C}_{ML}{}^+ A\boldsymbol{C}_{ML}\bar{X} = \bar{A}\bar{X} = \bar{Y}, \tag{24}$$

with $\bar{A}$ as under:

$$\bar{A} = \boldsymbol{C}_{ML}{}^+ A\boldsymbol{C}_{ML}. \tag{25}$$

In Eq. (25) $\boldsymbol{\mathcal{C}}_{ML}^{+} = C_L^{+}...C_1^{+}C_2^{+}C_3^{+}$ and $\boldsymbol{\mathcal{C}}_{ML} = C_3C_2C_1...C_L$. $\boldsymbol{\mathcal{C}}_{ML}^{+}$ can be computed as long as the elemental convolution maps, $C_i$ are invertible in a generalized sense. For example, we can combine POD and sGP maps to form a two-layer convolution, termed sGP-POD. The reverse combination is also possible, but not explored in this study. The multilayer convolution can also include transfer functions to modulate the evolution in the feature space by introducing desired nonlinearity (see Rowley and Dawson [11]). EDMD [20, 11] can be viewed as a combination of POD with a transfer function (TF) to build an extended polynomial basis up to a desired order. To restrict the dimensionality of EDMD, Williams et al. [21] included another POD layer for a POD-TF-POD mapping and implemented this through a kernel principal component analysis framework. The success of these different methods indicate that deep convolution could offer attractive improvements in performance.

### 3.5. Sparse Representation

Dimension reduction is a key component of building convolution maps so that a finite-dimensional approximation of the Koopman operator and a computationally tractable Markov model can be realized. POD-convolution provides a natural sparsification in basis space by optimizing capture of energy for the data and ordering the modes according to their energy content. Thus, truncating of POD basis with negligible energy content offers a natural reduction process. The success of this technique relies on the system being low-dimensional in the POD space and the low energy modes contributing negligibly to the dynamics. Often, DMD which uses POD convolution tries to fit a linear model for the physics using arbitrary number of basis that can lead to overfitting. Chen et al. [33] employs global optimization to build a robust DMD model from an arbitrary user-specified basis rank for a given data-set. In a similar vein, Wynn et al. [34] build a DMD model from an optimal orthogonal subspace of a given rank iteratively using matrix manifold theory [35, 36, 37].

The sparse Gaussian Process (sGP) convolution employs sparse representation in physical space by coupling a measurement matrix with a feature map embedded in a kernel. This approach can be viewed akin to building a on-demand model from sparse sensor data for online decision making. Given that the model fidelity is tied to the kernel hyper parameters and the sensor locations, we explore three different strategies for sparse sensor placement in this study:

**i.** pure k-means clustering over multiple partitions;

**ii.** combine k-means with centers carrying the most flow variability over time;

**iii.** place centers that minimize linear dependency in feature space using the SOGP algorithm [32].

For ease of identification, we adopt the following terminology: sGP-convolution with centers learned from k-means clustering in multiple partitions as sGP-k, centers learned from k-means clustering with variance measure as sGP-kv, and centers learned from sparse online GP algorithm as sGP-sv. K-means is a well-known clustering algorithm that partitions data (possible super set of sensor locations) into $k$ clusters, each associated with its own center that are chosen as the sparse sensor locations. This approach is system state independent and depends only on the initial super set of sensor location candidates, i.e. grid point locations in CFD data or pixel locations in PIV. K-means is known to be biased towards initial sampling density and to mitigate this, we partition the domain and assign a predetermined number of centers to be allocated to each such partition, i.e., k-means clustering in multiple partitions. For the cylinder wake flow the most interesting vortex shedding dynamics occur downstream of the cylinder. To capture this, we place 300 centers as

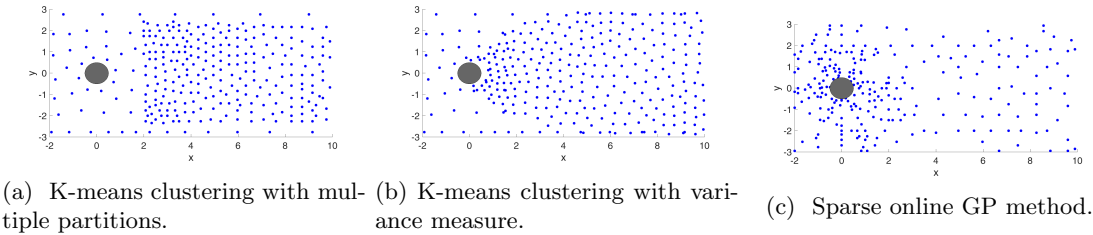

(a) K-means clustering with multiple partitions.  (b) K-means clustering with variance measure.  (c) Sparse online GP method.

Figure 1: Centers (sensor) placement learned from the different algorithms cylinder flow at $Re = 100$. 300 centers are chosen for this case.

shown in Fig. 1a by dividing the domain into three regions: the wake of the cylinder (250 centers), around the cylinder (30), and top and bottom (20). To introduce more dependence on the flow state in a systematic manner, we combine k-means centers with centers carrying the most temporal variability in the data snapshots (sGP-kv). As shown in Fig. 1b, we place 300 sensors (centers) equally split between the two different methods. Sparse Online Gaussian Processes or SOGP [32] is a purely data-driven Bayesian inferencing algorithm that identifies sensor (centers) locations based on incremental knowledge of the flow field streaming in so as to capture the essential dynamics of the system in the feature space. In this study, we adopt the implementation of the SOGP as available here in [38]. Figure 1c shows an example of centers computed from the SOGP algorithm using 30 random snapshots of the training data. In the following section, we will assess the effectiveness of these center placement algorithms on long-time prediction errors and ability to approximate the Koopman operator.

## 4. Numerical Experiments

To evaluate the different choice of convolution mappings, we consider three classes of canonical fluid flow problems: (a) unsteady cylinder flow with periodic vortex shedding (limit cycle attractor), (b) unsteady cylinder flow with transient vortex shedding (approaching limit cycle attractor), and (c) lock exchange problem (transient mixing problem). After briefly summarizing the data generation processes for these flow phenomena in subsection 4.1, we will perform Koopman spectral analysis on the limit-cycle flow past a cylinder at $Re = 100$ and $Re = 1000$ using the different convolution mappings in subsection 4.2. Finally, we will assess the effectiveness of multilayer convolution mappings for long-time model prediction for the various canonical fluid flows in subsection 4.3.

### 4.1. Data generation

#### 4.1.1. Cylinder Wake Flow

Flow past a cylinder [39, 40, 41, 11] has attracted considerable interest from the dynamical systems community for it particularly rich flow physics content, encompassing many of the complexities of nonlinear dynamical systems, while easy to simulate accuraetly on the computer using established CFD tools. In this study, we analyze the data at two separate temporal regions of interest: the periodic phase (limit cycle) and the transient phase (evolution towards limit cycle). To generate two-dimensional cylinder flow data, we adopt the spectral Galerkin method [42] to solve incompressible

Naiver-Stokes equations, as shown in Eq. (26), in our simulations.

$$\frac{\partial u}{\partial x} + \frac{\partial u}{\partial y} = 0, \tag{26a}$$

$$\frac{\partial u}{\partial t} + u\frac{\partial u}{\partial x} + v\frac{\partial u}{\partial y} = -\frac{\partial P}{\partial x} + \nu\nabla^2 u, \tag{26b}$$

$$\frac{\partial v}{\partial t} + u\frac{\partial v}{\partial x} + v\frac{\partial v}{\partial y} = -\frac{\partial P}{\partial y} + \nu\nabla^2 v, \tag{26c}$$

where $u$ and $v$ are horizontal and vertical velocity components. $P$ is the pressure field, and $\nu$ is the fluid viscosity. The rectangular domain used for this flow problem is $-25D < x < 45D$ and $-20D < y < 20D$, where $D$ is the diameter of the cylinder. The mesh was designed to sufficiently resolve the thin shear layers near the surface of the cylinder and transit wake physics downstream. The computational method employed fourth order spectral expansions within each element in each direction. Separate meshes were designed for the low ($Re = 100$) and high Reynolds number ($Re = 1000$) simulations to main accuracy of the computations. The sampling rate for each snapshot output is chosen as $\Delta t = 0.2$ seconds.

4.1.2. Lock-exchange problem

To examine various sparse convolution-based models on a flow problem that does not evolve into a limit-cycle dynamics, we model an unsteady lock-exchange problem [43], also called Boussinesq flow [44, 45], which exhibits strong shear and Kelvin-Helmholtz instability phenomena driven by temperature gradients. This mixing type flow is highly convective and evolves dynamically over time such that the training data will always be sparse for the dynamics of interest. To generate the data, we solve the dimensionless form of the two-dimensional incompressible Boussinesq equations [43], as shown in Eq. (27) on a rectangular domain that is $0 < x < 8$ and $0 < y < 1$.

$$\frac{\partial u}{\partial x} + \frac{\partial u}{\partial y} = 0, \tag{27a}$$

$$\frac{\partial u}{\partial t} + u\frac{\partial u}{\partial x} + v\frac{\partial u}{\partial y} = -\frac{\partial P}{\partial x} + \frac{1}{Re}\nabla^2 u, \tag{27b}$$

$$\frac{\partial v}{\partial t} + u\frac{\partial v}{\partial x} + v\frac{\partial v}{\partial y} = -\frac{\partial P}{\partial y} + \frac{1}{Re}\nabla^2 v + Ri\theta, \tag{27c}$$

$$\frac{\partial \theta}{\partial t} + u\frac{\partial \theta}{\partial x} + v\frac{\partial \theta}{\partial y} = \frac{1}{RePr}\nabla^2 \theta, \tag{27d}$$

where $u$, $v$, and $\theta$ are the horizontal, vertical velocity, and temperature, respectively. The dimensionless parameters $Re$, $Ri$, and $Pr$ are the Reynolds, Richardson and Prandtl numbers, respectively. In our simulation, we use $Re = 1000$, $Ri = 4.0$, and $Pr = 1.0$ and grid size of 1024 by 128. Initially fluids at two different temperatures are separated vertically at $x = 4$. The surrounding walls are adiabatic and have no-slip boundary condition. To obtain an accurate solution, a sixth-order compact finite difference scheme for the derivative approximations and third order Runge-Kutta time integration is adopted in Eq. (27).

## 4.2. Approximating the Koopman operator

To show that the sparse convolution framework can accurately capture the tuples of the Koopman operator for the two classes of convolution mappings presented in this article we model the limit cycle regime of the cylinder wake at two different Reynolds numbers, $Re = 100$ and $Re = 1000$. Since sych a system involves only stable or damped modes, it turns out to be a good test to ensure the convolution-based Markov models capture the correct spectral behavior. We know from research articles [6, 27] that DMD and consequently, the POD convolution framework captures a subset of the Koopman modes. In this analysis, we will compare the Koopman modes learnt using the sparse sGP-convolution for different choices of center locations with those from the POD-convolution that is representative of the Koopman operator for a system that evolves on an attractor. The associated eigenvalues are shown in Figs. 2a and 2b in the form of real and imaginary parts plotted with a unit circle for the two different cylinder wakes at $Re = 100$ and $Re = 1000$. Figs. 2a and 2b figures clearly show that all the different sGP-convolutions capture the leading eigenvalues of the Koopman operator. For the case of $Re = 100$, the eigenvalues located at the right plane of the Ritz plot are accurately captured for the sGP-convolutions as compared to the POD-convolution. For the case of $Re = 1000$, the ones that reside close to the unit circle from sGP-convolutions share similar identities to the ones from POD-convolution in Fig. 2b. However, there does exist some mismatch for the eigenvalues that reside on the left plane close to the unit circle. To provide further quantification, we also tabulate the first three dominant eigenvalues in Table 1. In comparison, we observe that the first three leading eigenvalues are accurately extracted among all four approaches irrespective of the choice of sensor placement algorithm employed for the sGP case for both $Re = 100$ and $Re = 1000$. The agreement was observed for other lesser dominant eigenvalues as well. In account of the connection between the POD-convolution (DMD) and Koopman modes, we can say that the sGP-convolution also approximates the Koopman operator. Fig. 3 shows the three dominant eigenmodes for the models generated using different convolution operators for $Re = 100$. It is worth noting that the eigenmodes computed directly from the sGP-convolution produce aliasing or deconvolution errors as the convolution operator $C$ is not exactly invertible. This aliasing effect can be minimized by choosing the sensor locations optimally and increasing their number. While there have been efforts to overcomes these deconvolution errors in the context of DMD by projecting the eigenmodes of the reduced operator onto the full uncompressed state vector instead of the sparse state vector [46] to obtain the Koopman modes, this discussion is outside the scope of this article. Out of all three sGP-convolutions, the case (sGP-kv-convolution) with centers learned from k-means with variance measure produces the best eigenmodes. This is possibly from the placement of sufficient sensors in the wake region of the flow (Fig. 1b) which helps with improved prediction. During the deconvolution procedure, a Tikhonov regularization is employed make the solution unique. This can be viewed as equivalent to the SVD truncation applied in the case of POD-convolution.

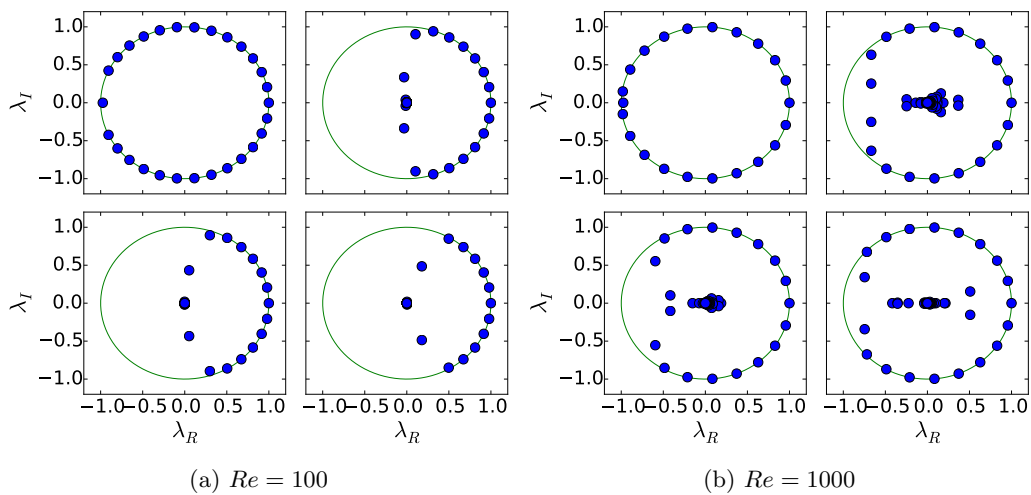

(a) $Re = 100$  (b) $Re = 1000$

Figure 2: Eigenvalues computed from $\bar{A}$ are shown as a Ritz plot for the various convoltion models: POD-convolution(top left), sGP-k-convolution(top right), sGP-kv-convolution(bottom left),and sGP-sv-convolution(bottom right) for the periodic cylinder flow at $Re = 100$ (left) and $Re = 1000$ (right).

Table 1: The first three dominant eigenvalues of the Koopman operator extracted from POD-convolution(POD), sGP-k-convolution, sGP-kv-convolution, and sGP-sv-convolution for periodic cylinder flows at $Re = 100$ and $Re = 1000$.

| | $Re = 100$ | | |
|---|---|---|---|
| Convolution | Eigenvalue 1 | Eigenvalue 2 | Eigenvalue 3 |
| | $f = 0.1655$ | $f = 0.3311$ | $f = 0.4966$ |
| POD | $0.9784 + 0.2065i$ | $0.9147 + 0.4041i$ | $0.8115 + 0.5843i$ |
| sGP-k | $0.9784 + 0.2065i$ | $0.9146 + 0.4041i$ | $0.8113 + 0.5841i$ |
| sGP-kv | $0.9784 + 0.2065i$ | $0.9145 + 0.4040i$ | $0.8112 + 0.5841i$ |
| sGP-s | $0.9784 + 0.2065i$ | $0.9146 + 0.4041i$ | $0.8114 + 0.5842i$ |
| | $Re = 1000$ | | |
| Convolution | Eigenvalue 1 | Eigenvalue 2 | Eigenvalue 3 |
| | $f = 0.2367$ | $f = 0.3311$ | $f = 0.4966$ |
| POD | $0.9561 + 0.2930i$ | $0.8283 + 0.5603i$ | $0.6277 + 0.7785i$ |
| sGP-k | $0.9561 + 0.2930i$ | $0.8282 + 0.5604i$ | $0.6277 + 0.7784i$ |
| sGP-kv | $0.9561 + 0.2930i$ | $0.8282 + 0.5603i$ | $0.6277 + 0.7785i$ |
| sGP-s | $0.9561 + 0.2930i$ | $0.8282 + 0.5603i$ | $0.6277 + 0.7785i$ |

### 4.3. Model prediction

In the previous subsection, we quantified the effectiveness of the sparse convolution framework for Koopman spectral analysis. In this section, we assess the model performance to predict the evolution of nonlinear fluid flows in time beyond the training region. For this purpose, we consider three different classes of temporally evolving flow systems: (i) a periodic vortex shedding flow (limit cycle) past a cylinder at $Re = 100$; (ii) a transient wake instability behind a cylinder at $Re = 100$ that evolves into a periodic vortex shedding dynamics and (iii) a dynamically evolving buoyancy-driven Boussinesq mixing layer ($Re = 1000$) without any repetitive behavior. In all these cases, a certain amount of training data is chosen to learn the underlying Markov model the feature space which is then used to predict the evolution of the dynamics.

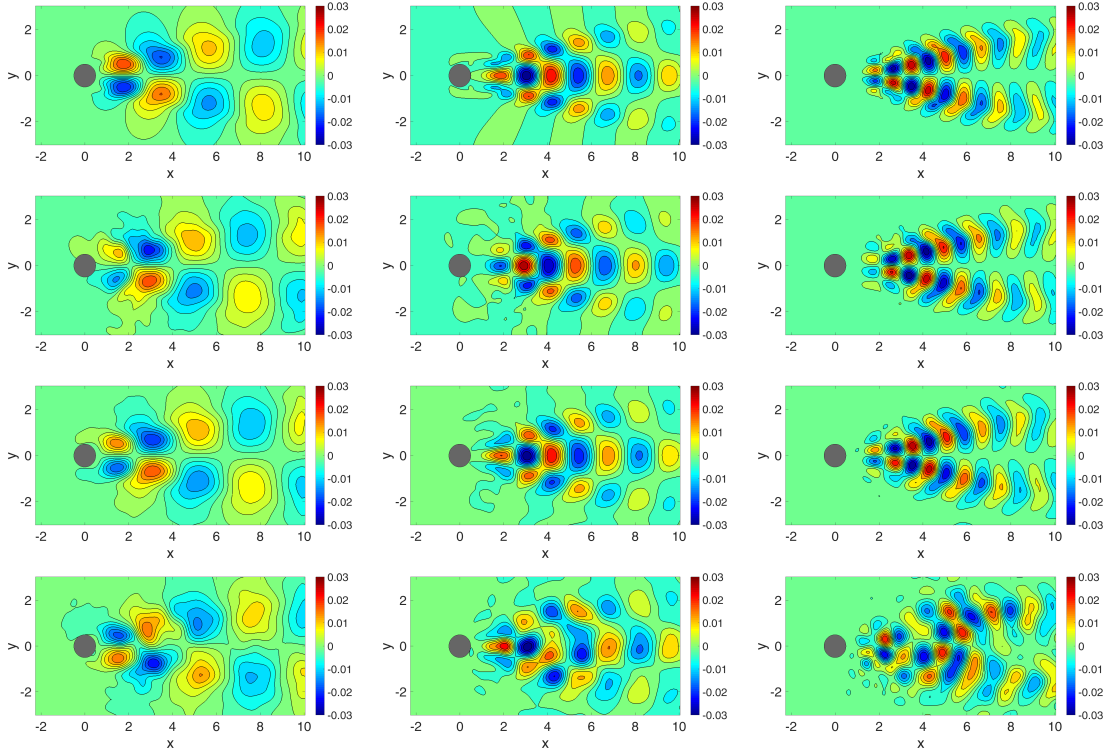

Figure 3: Comparison of the first three Koopman eigenmodes for the different convolution mappings are shown for $Re = 100$ :. The left figures correspond to the first dominant eigenmode, the middle is the second dominant and the right is the third dominant mode. From top to bottom: POD-convolution, sGP-convolution, sGP-kv-convolution, and sGP-sv-convolution.

#### 4.3.1. Prediction of Limit-cycle Dynamics

The flow in the wake of cylinder with periodic vortex shedding has been a successful case study for reduced order models due to its predictable limit-cycle dynamics. In this study, we choose 300 snapshots of data corresponding to a non-dimensional time $(T = \frac{Ut}{D})$ of $T = 60$ with uniform spacing of $dT = 0.2$. This corresponds to about ten cycles of periodic vortex shedding behavior. We then use the trained models to predict up to $T = 400$ which amounts to 2000 snapshots. We compare the predictive performance for both the POD- and the sparse GP class of convolution models with different sensor-placement algorithms. The limit-cycle behavior is observed in both the POD-convolution (Fig. 5) as well as the sGP convolution with k-means clustering & variance measure algorithm (sGP-kv) for placement of 300 centers (Fig. 4). The left plots in these figures show early time predictions while the right shows the later predictions. The other sensor placement algorithms produce models with qualitatively similar predictive performance. We quantify the predictive performance of the various models using $L^2$ error norms for the prediction of the weights:

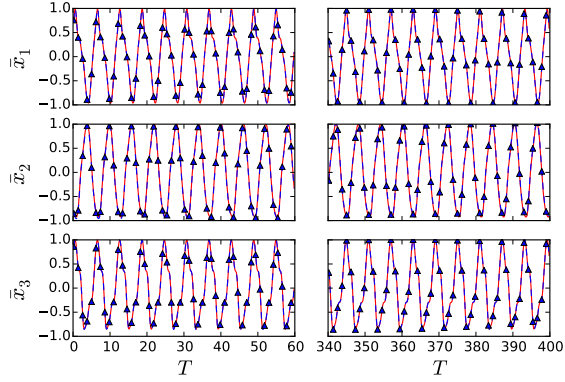

Figure 4: The first three projected (red line) and predicted (blue triangle) weights for the periodic cylinder flow at $Re = 100$ using sGP-kv-convolution with 300 centers. The left figures show early and the right, later time predictions.

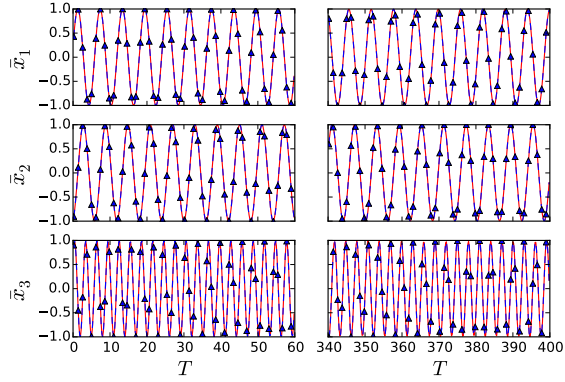

Figure 5: The first three projected (red line) and predicted (blue triangle) weights for the periodic cylinder flow at $Re = 100$ using POD-convolution. The left figures show early and the right, later time predictions.

$$\phi_j = \sqrt{\frac{1}{K}\sum_{k=1}^{K}(\bar{x}_{k,j}^T - \bar{x}_{k,j}^P)^2} \quad . \tag{28}$$

Similarly, we also quantify the predictive performance of the different sparse convolution mappings using $L^2$ error norms for the rebuilt solution state as:

$$\psi_j = \sqrt{\frac{1}{N}\sum_{i=1}^{N}(x_{i,j}^T - x_{i,j}^P)^2} \quad , \tag{29}$$

where $i$ and $j$ represent the state vector dimension and snapshot identity of the data, respectively. From Eq. (9), the projected weights, denoted as $\bar{x}_{k,j}^T$, can be computed by projecting the solution state onto the convolution operator, whereas the predicted weights, $\bar{x}_{k,j}^P$ are obtained from the trained model. $K$ and $M$ represent the total number of weights and snapshots, respectively. Table 2 and 3 summarize the computed error metric from our simulations after sixty non-dimensional times ($\phi_{T=60}$, $\psi_{T=60}$) and 400 non-dimensional times ($\phi_{T=400}$, and $\psi_{T=400}$) respectively. The corresponding time series of the $L^2$ error norms for $\phi_j$ and $\psi_j$ are shown in Fig. 6 and Fig. 7, respectively. The tabulated error metrics (Table 2 and Table 3) show that for the prediction of the weights, the sGP convolution is preferred, but for the prediction of the flow field which include the deconvolution step, the POD-convolution model is the most accurate. The deconvolution errors in the sGP methods impact the full field solution reconstruction. Further, for the various sGP methods, the sensor placement has little impact, but the number of sensors matter as shown in Table 3. To assess, the prediction performance using as a multilayer convolution we layer a POD-convolution over the sGP convolution (sGP-kv) for a sGP-kv-POD mapping. While this approach is the most efficient method for learning the model, it also seems to have an order of magnitude smaller error in the weight predictions as compared to the regular single layer sGP models. However, the prediction error in the solution field matches that of the a single layer sGP-kv model as expected. A consistent theme observed for all the models is that the prediction error continues to grow with time irrespective of the algorithm employed as shown in Fig. 7. From Fig. 8, it shows sample instantaneous reconstruction of the solution field predicted using the various sGP-convolution models at $T = 400$. These images clearly show the manifestation of the deconvolution errors. However, there are also visible contributions from lack of sensors near the edges of the simulation domain which might skew the error metrics. Since, the prediction region of interest is the wake of the cylinder flow with the vortex shedding dynamics, we compute the solution reconstruction error for the wake region in isolation. These error metrics are plotted over time in Fig. 9 and confirm that the most accurate sensor placement is using the sGP-kv framework for this fluid flow problem. As the number of centers used in sGP-convolutions is increased to 600, both the prediction of the weights and field prediction improve notably as shown in Fig. 6 and 7. The reconstructed solution from the sparse sensors is also observed to improve significantly. The above findings indicate that the number and the location of sensor placement impacts the model prediction performance for the sGP convolutions. Ideally, we want to place centers at every location where the flow information is available to capture the system dynamics, but the cost of model building would be too expensive. This sGP convolution framework is explicitly targeted towards experimental set-ups where data is often obtained from sparse sensors and online predictive models need to be built for real-time control. Consequently, the problem of sensor placement evolving as data streams is very relevant.

Table 2: The $L^2$ error metrics for the prediction of the weights and solution field using different convolutions (sGP with 300 centers) for $Re = 100$ periodic cylinder flow. sGP-kv-POD convolution retain 100 modes. POD-convolution retain 10 modes which 99.99 percent of energy are captured.

| Convolution | $\phi_{T=60}$ | $\psi_{T=60}$ | $\phi_{T=400}$ | $\psi_{T=400}$ |
|---|---|---|---|---|
| sGP-k | 4.9562E-3 | 1.2920E-1 | 1.1207E-2 | 1.2970E-1 |
| sGP-kv | 5.4338E-3 | 9.9607E-2 | 2.1339E-2 | 9.9893E-2 |
| sGP-sv | 7.7917E-3 | 4.7397E-2 | 2.2470E-2 | 5.1750E-2 |
| sGP-kv-POD | 3.5863E-4 | 9.9604E-2 | 1.5987E-3 | 9.9894E-2 |
| POD | 2.7619E-2 | 6.2501E-4 | 1.2729E-1 | 1.4183E-3 |

Table 3: $L^2$ error metrics for the prediction of the weights and solution field from different convolutions (sGP with 600 centers) for $Re = 100$ periodic cylinder flow.

| Convolution | $\phi_{T=60}$ | $\psi_{T=60}$ | $\phi_{T=400}$ | $\psi_{T=400}$ |
|---|---|---|---|---|
| sGP-k | 1.6181E-3 | 7.0818E-2 | 4.2919E-3 | 7.0842E-2 |
| sGP-kv | 3.4440E-3 | 5.1194E-2 | 8.3614E-3 | 5.1272E-2 |
| sGP-sv | 2.5959E-3 | 2.7470E-2 | 6.6308E-3 | 2.7593E-2 |

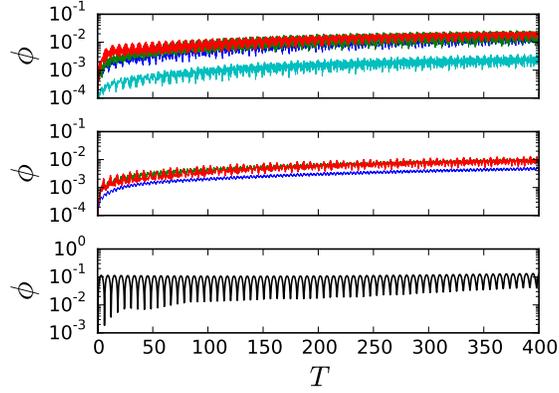

Figure 6: $L^2$ error norms of the weights prediction field for periodic cylinder flow at $Re = 100$. Blue: sGP-k-convolution. Green: sGP-kv-convolution. Red: sGP-sv-convolution. Cyan: sGP-kv-POD-convolution. Black: POD-convolution. Top: sGP convolutions with 300 centers. Middle: sGP convolutions with 600 centers.

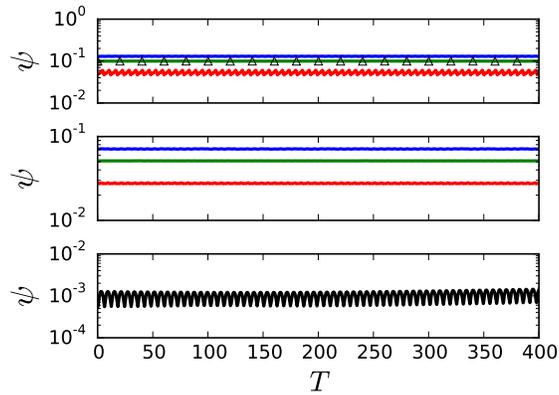

Figure 7: $L^2$ error norms of the solution field for periodic cylinder flow at $Re = 100$. Blue: sGP-k-convolution. Green: sGP-kv-convolution. Red: sGP-sv-convolution. Triangle: sGP-kv-POD-convolution. Black: POD-convolution. Top: sGP convolutions with 300 centers. Middle: sGP convolutions with 600 centers.

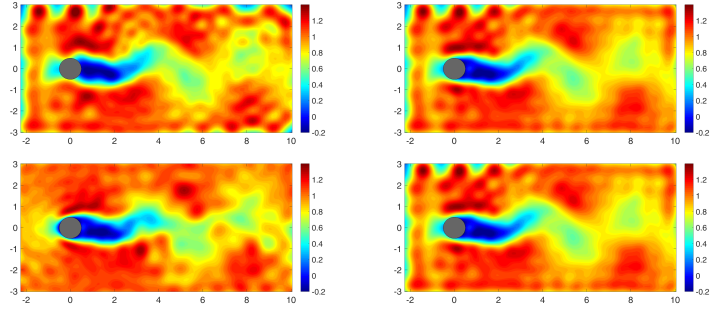

Figure 8: The stream-wise velocity contour of the predicted solution field at $T = 400$ for the periodic cylinder flow at $Re = 100$ from sGP-convolutions with 300 centers. Top left: sGP-k-convolution. Top right: sGP-kv-convolution. Bottom left: sGP-sv-convolution. Bottom right: sGP-sv-POD-convolution.

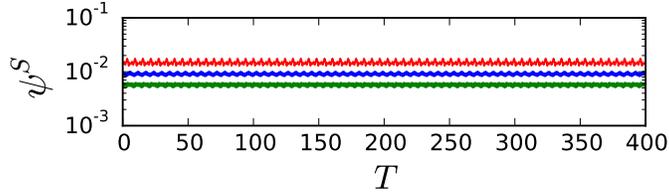

Figure 9: $L^2$ error norms of the solution field for selected domain for periodic cylinder flow at $Re = 100$. Blue: sGP-k-convolution. Green: sGP-kv-convolution. Red: sGP-sv-convolution. sGP-convolutions performed in this figure use 600 centers.

4.3.2. Prediction of Transient Nonlinear Dynamics Evolving towards Limit-cycle

The previous section focused on the predictive accuracy of the sparse convolution models to predict the evolution of a limit-cycle behavior over long times. This study showed a very gradual, but small accumulation of predictive error over long time periods for the different sparse convolution models which can be controlled to an extent with regularization. In this section, we explore the ability of the data-driven models to predict temporally evolving dynamics over long-time using limited amount of training data. It is well known that temporally evolving systems are data-sparse by construction while building data-driven models. For this analysis we chose to train the data over the cylinder wake flow regime when it just about becomes unstable and evolves into a limit-cycle periodic vortex shedding. Like the earlier study, we use a small number of snapshots (340 corresponding to $T = 68$) of data from the transient unstable regime such that the first half of the data lies in the transient phase, and the second half lies at the limit-cycle regime. The analysis of predictive performance is carried out for the various sparse convolution frameworks including sGP (sGP-k, sGP-kv and sGP-sv with 600 sensor locations) and POD-convolution (DMD). In addition, we also assess the performance of multilayer convolution ideas of the following architectures: sGP-kv-POD, POD-TF, POD-PolyK convolutions. Here, TF represents an extended polynomial basis to represent the second layer of nonlinear mapping from the POD weights to another feature space as is done in EDMD [20] and PolyK represents the kernel trick implementation of this two-layer model as in Williams et al. [21]. The POD-PolyK which is kernel DMD method is essentially a POD-convolution with a kernel trick layered on top of EDMD (POD-TF) so that the larger dimension of features in the polynomial basis space can be controlled. In all these cases, the prediction from the model is carried out to nearly three times the training region size for a non-dimensional duration of $T = 200$ (1000 snapshots).

The $L^2$ error metric for the prediction of weights ($\phi$) and solution field ($\psi$) for all the seven different cases are shown in Figs. 10 and 11, respectively. The first observation here is that the plain single-layer POD-convolution or DMD does predict this transient nonlinear dynamics as seen from the dynamical evolution of the projected and predicted weights shown in Fig. 12a although it represents the convolution operator built using the most compact data-driven basis for this flow phenomena. To assess the role played by the choice of POD basis used in convolution operator, we perform SVD on the periodic regime instead of computing the SVD on the entire training data, which is then used to build $C$ and resulting linear Markov model. As seen from Fig. 10, this modification in $C$ does not seem to improve the prediction accuracy much indicating that the problem lies in the linear model $\bar{A}$ not being able to represent the dynamics in feature space. In fact, this result is consistent with the results shown in Rowley and Dawson [11] for such transient nonlinear dynamics. To address this deficiency, Williams et al. [21, 20] and Rowley and Dawson [11] employ nonlinear basis such as high order polynomials in EDMD and its kernel variant to map into a feature space so one can capture the nonlinear evolution of the POD coefficients. The equivalent algorithms in the sparse multilayer convolution terminology are POD-TF and POD-PolyK respectively, i.e., layering a nonlinear transfer function and a polynomial kernel over a POD-convolution. As observed from Figs. 10 and 11, both these extended basis approaches produce the least error metric over time for the prediction of both the weights in feature space and full state vector in input space. The corresponding predicted and projected weight trajectories for the POD-TF framework are shown in Fig. 12b. The reconstructed full state for the POD-TF is shown in Fig. 13b which is qualitatively similar to the actual solution (Fig. 13a). In contrast, the solution reconstruction for the single-layer POD-convolution (DMD) method (Fig. 13c) shows highly delayed growth of the wake instability.

On the other hand, the sGP-convolution models capture the correct trajectory of the weights in the feature space over long-times in spite of being a single-layer convolution framework with no explicit nonlinearity embedded into the convolution. As seen in Fig. 10, the error in the weights prediction is comparable to that using the multilayer mappings with extended nonlinear basis functions such as POD-TF (EDMD). The predicted and the projected weights using sGP-kv sensor placement is shown to match each other in Fig. 12. However, as expected, the error for the reconstructed solution (Fig. 11) is much higher than that for the POD-TF (EDMD) and POD-PolyK (Kernel DMD) due to the errors from inexact deconvolution from the sparse space representation. The visualization of the reconstructed solution for both the sensor placement strategies (Figs. 13d and 13e) show the manifestation of these errors, but surprisingly predict the overall vortex breakdown dynamics correctly irrespective of the sensor placement strategy adopted. The effect of sensor quantity on the predictive accuracy for such flow dynamics is not explored here and will be considered as future work.

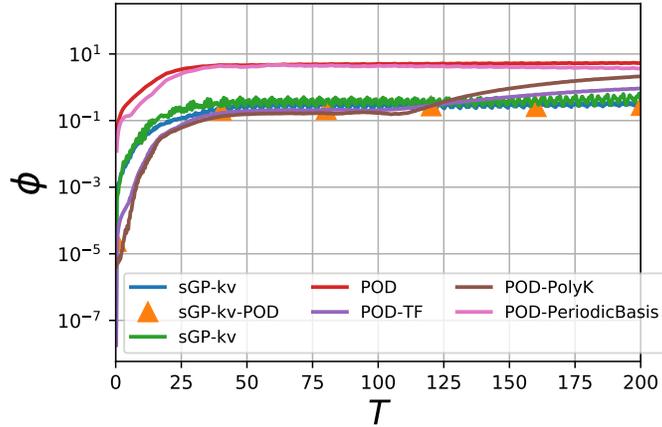

Figure 10: $L^2$ error norms of the predicted weights for transient cylinder flow at $Re = 100$.

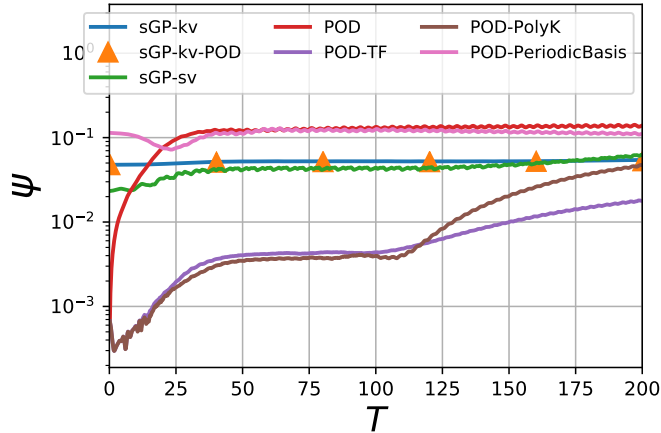

Figure 11: $L^2$ error norms of the solution field for transient cylinder flow at $Re = 100$.

The data-specific characteristic of the sGP-convolution is primarily embedded in the measurement matrix carrying the sensor locations and the kernel hyperparameters. While the sGP convolution operator is designed specific to the training data set, it is able to accommodate the dynamically evolving physics due to the fine-scale granularity of information contained in the numerous features (600 in the case). That is, the nonlinear evolution of the physics is distributed among the many features and less concentrated. In the case of POD-convolution, since the features are low-dimensional (10 modes) due to an optimally sparse mapping, the transient dynamics (from growth of instability to limit-cycle) is represented using a small number of features that make it hard to approximate using a simple linear operator. This is a hypothesis that is yet to be verified and currently explored. A potential criticism of the sGP methods is the increased computational cost due to the use of more features as compared to the POD-convolution. To address this, we considered a sGP-kv-POD multilayer convolution model that reduces the complexity of the sGP features (600 kernels) by performing a POD convolution to a size of 100 modes. The prediction error metrics for this case (Figs. 10 and 11) indicate that the added convolution layer has minimal impact on the accuracy.

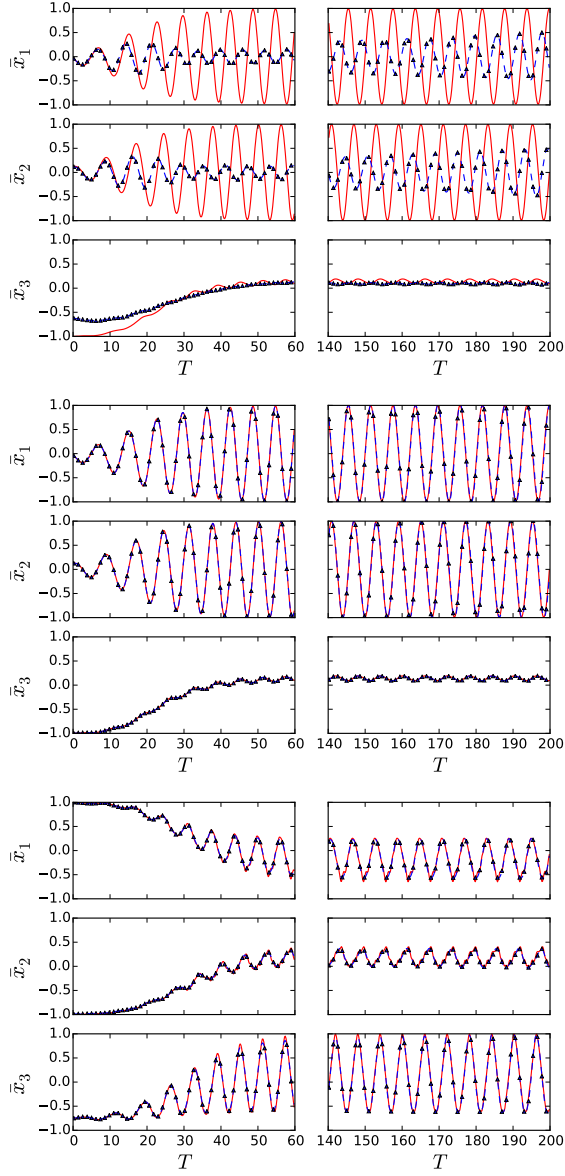

Figure 12: The evolution over time of the first three projected (red) and predicted (blue) weights (in feature space) for the transient cylinder flow at $Re = 100$ using POD-convolution (top), multilayer POD-TF-convolution (middle), and sGP-convolution using sGP-kv (bottom). The left column in each figure show early-time predictions and the right column shows long time behavior.

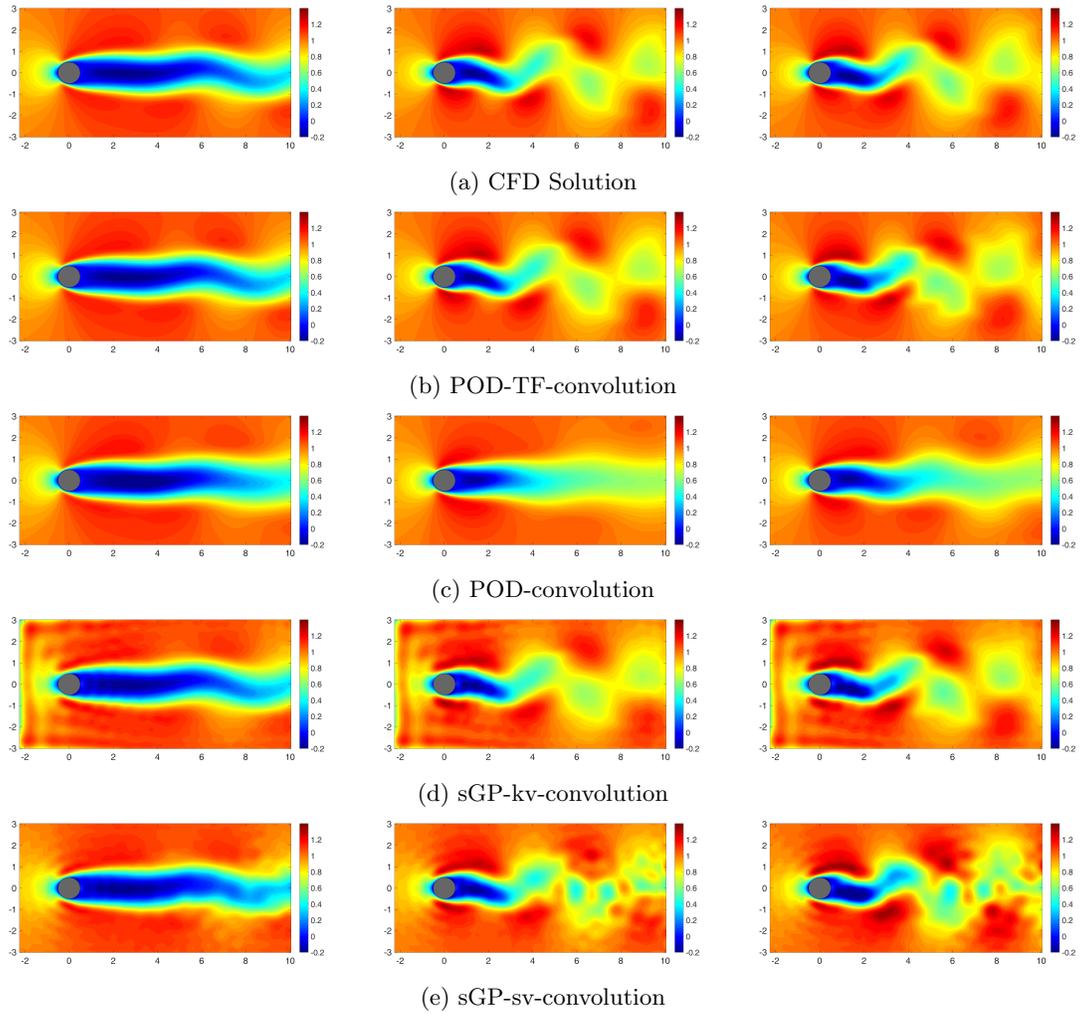

Figure 13: Comparison of the predicted solution for three different instances in time: $T = 25$, $T = 68$ and $T = 200$ for the different sparse convolution models with the corresponding solution using spectral Galerkin (CFD) methodology.

### 4.3.3. Prediction of Dynamically Evolving Buoyant Boussinesq Mixing Flow

In the previous two case studies, we focused on the physics that either evolved on a limit-cycle or approached limit-cycle behavior and are relatively, low-dimensional systems. Data-driven methods in general tend to perform reasonably well as long as the underlying physics of the flow does not vary significantly and convolution maps remain relevant to the evolving input space. If not, long-time predictions of such systems will be challenging and prone to errors. In this section, we study a 2D buoyant Boussinesq mixing flow or commonly, *the lock exchange problem* [43, 44]. The physics here does not lend itself to limit-cycle behavior and evolves through convective and shear instabilities (Kelvin-Helmhotlz). San and Borggaard [43] show that such flow physics require dynamically updated choice of POD-basis for prediction using Galerkin projection on the flow equations. In this analysis, we compare the predictive model performance for a reduced suite of sparse convolution models including POD, POD-TF, sGP-kv and sGP-kv-POD for a total of two single-layer and two multilayer convolutions. For the training set, we use 400 snapshots ($T = 8s$) of data and attempt to predict up to 800 snapshots ($T = 16s$). As was observed for the transient cylinder wake flow, the POD-convolution produces the most error for the prediction of the weights (Fig 14) while the multilayer POD-TF (EDMD) convolution provides the least prediction error within the training region. The POD-convolution framework is unable to capture the quadratic nonlinear evolution of the weights using a linear model. However, both the single and multilayer sGP-convolution models perform much better in capturing this nonlinear dynamics within the training region $T = 0 - 8s$. This is consistent with earlier observations for the transient cylinder flow where the sGP in spite of being a linear convolution framework, is able to predict the transient nonlinear dynamics from $T = 0 - 8s$ as shown in Fig. 16. Clearly, the predicted weights (blue triangles) mimic the projected weights (red curve) in the first half of this plot. As expected, the multilayer POD-TF convolution (EDMD) which embeds quadratic nonlinearity explicitly in the convolution map predicts the weight evolution in feature space accurately within the training region (Fig. 17). The full flow state reconstruction error in the training region (Fig. 15) clearly shows that the POD-TF convolution performs best amongst that various candidate mappings whereas the sGP predicts the weights accurately, but introduces reconstruction errors from the deconvolution step.

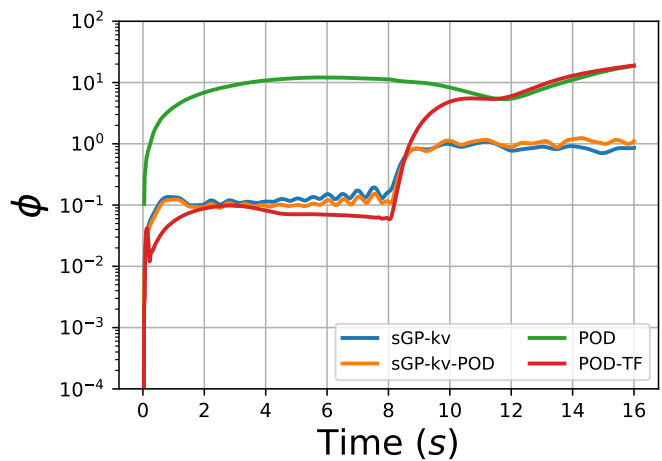

Figure 14: $L^2$ error norms of the predicted weights for the lock-exchange problem

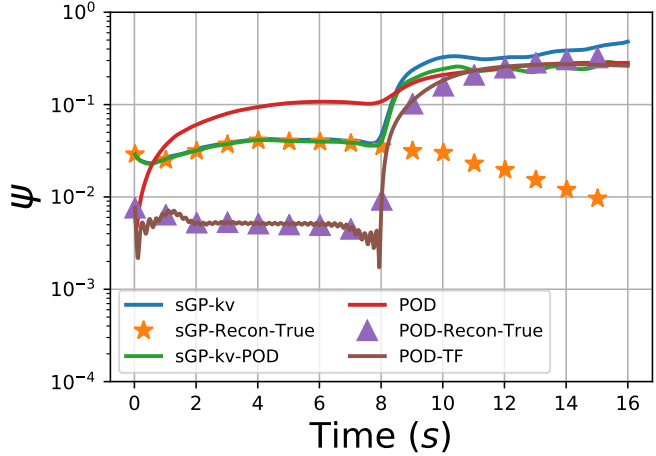

Figure 15: $L^2$ error norms of the solution field (pred - true) for the lock-exchange problem.

Beyond the training region, all the predictive models display a jump in error both in the weights and full field predictions (Figs. 14 and 15) after $T = 8s$. Both the sGP and the POD-TF convolution models that performed well in the training region see increased errors in the true prediction region. This is attributed to the fact that the evolution of the projected weights occur differently beyond the training region due to the nonlinear dynamics and is not captured by any of the single and multilayer sparse convolution models (see Figs. 16 and 17). In fact, the POD-TF comes closest to predicting the evolution of the weights in the prediction region (Figs. 17) due to the embedded quadratic nonlinearity, but is still not accurate enough. On the other hand, sGP model predictions beyond $T = 8$ are grossly inaccurate (Fig. 16). The prediction error in the full field reconstruction is a combination of the prediction errors in the feature space and the deconvolution (reverse mapping) errors. The above discussion leads to the following question: How accurately would the sparse convolution models predict the full field state if the evolution of the features were predicted accurately? To answer this, we performed deconvolution on the exact projected weights (the red curves in Figs. 16 and 17). The error metric for this exact reconstructed solution (i.e. difference between reconstructed and true solutions) is shown in Fig. 15 as the curves named "POD-Recon-True" and "sGP-Recon-True" for the POD- and sGP-convolution mappings, respectively. It is clearly seen that the instantaneous error for the full state reconstruction using the exact projected weights and model predicted weights is comparable for both the POD and sGP convolutions. Thus, the significant source of the error jump from the training to the prediction regions arise from the deconvolution step. In Fig. 18 we compare the actual temperature field (part (a)) at four different instances in time with the reconstructed solution using exact projected weights (part (b)) and model predictions (part (c)). This clearly illustrates the issue with the POD-convolution-based models. The reconstructed solution using exact projected weights (Fig. 18b) and predicted weights (Fig. 18c) matches the actual CFD simulation (Fig. 18a) at $T = 4, 8s$. However, at $T = 12, 16s$ (outside the training region) the reconstructed solution using exact weights looks nonphysical. This arises because the POD basis generated from the training data set and used to construct $C$ does not the span the subspace containing the actual solution outside the training region. Thus, the solution field cannot be optimally reconstructed from the predicted weight evolution in the feature space even if its is exact. Further, if one were to use the entire data (both training and prediction validation regimes) to build the POD basis, then the evolution of the features is quite different from that when one uses only the training data. This is shown as the green curve in Fig. 17.

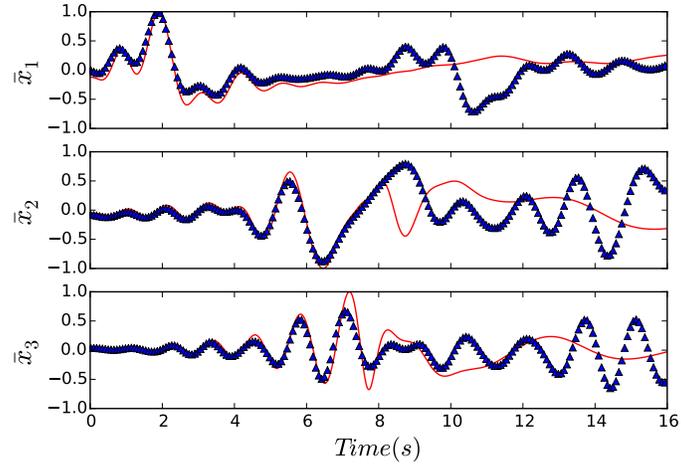

Figure 16: The first three weight evolution with time for the lock-exchange problem using sGP-kv-convolution. Red line: projected weights. Blue triangle: predicted weights.

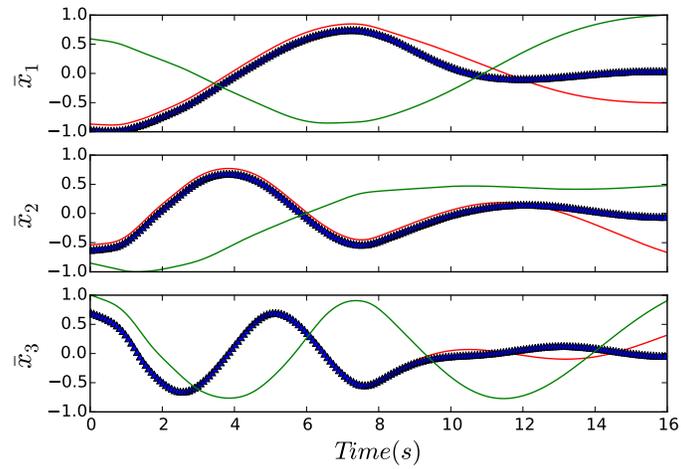

Figure 17: The first three weight evolution with time for the lock-exchange problem using POD-TF-convolution. Red line: projected weights. Green line: true weights using the entire data set to learn the POD basis for $C$. Blue Triangle: predicted weights.

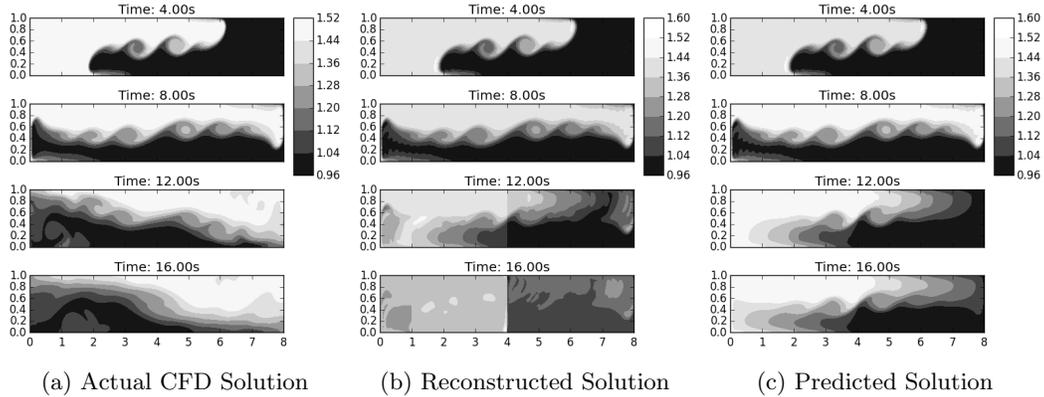

(a) Actual CFD Solution   (b) Reconstructed Solution   (c) Predicted Solution

Figure 18: Comparison of the solution (temperature) field for the buoyant-mixing flow predicted using multilayer POD-TF (EDMD) solution with actual CFD simulation at four different instances in time: $T = 4, 8, 12, 16$s split equally between the training and prediction regions of the simulation. The three sets of figures correspond to: (a) Actual CFD solution; (b) Reconstructed solution using exact projected weights; and (c) Reconstructed solution using predicted weights.

Fig. 19 shows the comparison of the actual field with the exact reconstructed and predicted solution fields for sGP-kv convolution framework. In the sGP-convolution models, the underlying kernels are generic and the data-specific information enters through the kernel hyperparameters including the sensor locations. As a result, the convolution operator $C$ also tends be more generic, especially if sufficient sensors can be placed as the sensitivity to the data is minimized. Fig. 19b shows that although the sGP-kv formulation produces noisy exact reconstructions of the projected weights due to the well-known deconvolution issues, they do capture the qualitatively correct physics evolution even beyond the training region. Of course, this requires accurate estimation of the evolution of the weights in the feature space. On the other hand, if the weight evolution is not accurate in the prediction region (see Fig. 16), then the reconstructed solution using the predicted weights is inaccurate as shown in Fig. 19c. The quantification of this observation is clearly seen in Fig. 15, where the "sGP-kv" predictions deviate from the "sGP-Recon-True" curves beyond the training region. In summary, the sGP-convolution with sufficient sensors is less data-dependent and the prediction quality is tied to the prediction accuracy of the weights in the feature space. The price one pays for such a generic sparse convolution is in dealing with a higher-dimensional feature space (1500 kernels in this case) which thankfully can be reduced by nearly an order-of-magnitude by layering a POD convolution on top of the sGP-kv operator. In Figs. 16 and 17), we see that the multilayer sGP-kv-POD convolution is as accurate as the more expensive single-layer sGP-kv convolution.

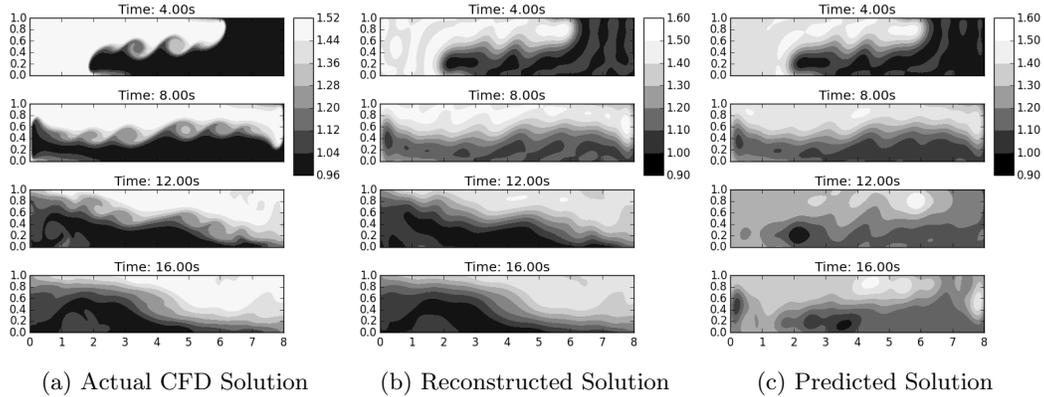

Figure 19: Comparison of the solution (temperature) field for the buoyant-mixing flow predicted using single-layer sGP-kv solution with actual CFD simulation at four different instances in time: $T = 4, 8, 12, 16$s split equally between the training and prediction regions of the simulation. The three sets of figures correspond to: (a) Actual CFD solution; (b) Reconstructed solution using exact projected weights; and (c) Reconstructed solution using predicted weights.

## 5. Conclusions

In this article, we explore and assess strategies for data-driven modeling of nonlinear fluid flows using sparse convolution-based mapping into a feature space where a linear Markov model describes the dynamics of the underlying system. In particular, we note that if the convolution map from the input and output states to the corresponding feature space are the same, then the resulting Markov process can be related to the class of Koopman operator theoretic methods and evolutionary kernel methods. Thus, in a unified view, popular techniques for Koopman operator approximation such as DMD, EDMD and the many successful evolutionary kernel methods such as EGP differ primarily in the choice of the sparse convolution map. To illustrate this, we show that the sGP-convolution-based linear Markov model can approximate the same Koopman tuples as predicted using DMD. In addition, we also explore a way to build complex convolution maps for modeling highly nonlinear and higher dimensional systems by layering multiple low-dimensional convolution operators as long as they are each independently invertible. This appears to have the potential to help bypass the curse of dimensionality encountered in EDMD-class of methods. The success of the sparse convolution models ultimately hinge on their ability to accomplish two things: (a) to map across the input and feature spaces accurately and efficiently and (b) to predict the dynamical evolution of the system in the feature space accurately and efficiently. This requires the basis constituting the convolution operator to accurately span the input states and the transformation map to be spanned by the Koopman eigenfunctions that are not known *a priori*. Given this, we assessed two prominent classes of convolution maps for their ability to predict long-term dynamics of nonlinear fluid flows with limited training data including a wake flow limit-cycle attractor, the transient evolution of a cylinder wake to a limit cycle attractor and finally, a highly transient buoyant mixing flow. The first is the class of purely data-driven feature maps based on singular value decomposition (SVD) or POD-convolution while the second is the more generic class of sparse kernel-embedded feature maps such as sGP convolution.

sGP-convolution maps incorporate sparse sensor placement whose locations impact long-term prediction errors. POD-convolution models represent optimal mapping between feature and input spaces, but their data-driven nature make them suceptible to error accumulation in modeling transient nonlinear cylinder wake dynamics. As in EDMD, embedding a nonlinear convolution layer with the POD-map is required for predicting the transient limit-cycle dynamics. The sGP convolution models, in spite of not including any explicit nonlinearity in the mapping and contending with deconvolution errors, were able to predict the nonlinear wake breakdown with reasonable accuracy. However, this success comes at a higher cost due to a higher-dimensional feature space in sGP-models. We show that this increased cost can be controlled by adding a POD-convolution layer on top of the sGP to build a multilayer convolution with lower dimensionality and similar prediction accuracy.

For a highly transient fluid flow dynamical system such as buoyant mixing, the data specific singular vectors in the POD-convolution become outdated. In spite of their optimal deconvolution properties and compact representation in feature space, such methods are deficient in predicting the dynamics beyond the training region. To make this work, the singular vectors need to be updated dynamically and is currently being explored by the authors. The generic kernel structure of sGP-convolutions allow more flexibility in handling transient flows in spite of their tendency to introduce deconvolution errors. Minimizing deconvolution errors require more sensors being placed judiciously to span highly evolving physics. In our results, we observe that the success of the sGP-based models for this buoyant mixing flow arises from the ability of the convolution operator to satisfactorily represent the flow state all through its dynamical evolution. In spite of the sensor placement being data-driven, we note that in the limit of a large number of centers (and consequently, more features), their locations become relatively insensitive to the underlying flow structures. Consequently, such methods tend to be successful as long as the evolution of the dynamics in the feature space is accurately modeled - a task easier said than done. In fact, for the buoyant mixing problem, the linear Markov model in the feature space is incapable of capturing the nonlinear evolution of the sGP weights and may require embedding additional nonlinear basis functions in the convolution operator. However, the overall flow state predictions show qualitatively accurate characteristics. As the features grow with the number of centers in the sGP models, a multilayer convolution strategy with POD-based dimensional reduction appears highly effective with minimal loss of accuracy.

In conclusion, sparse-convolution-based Markov linear models require projection onto a basis that can evolve with the system so that it can optimally represent its state at any given instant. Multilayer convolution has shown success based on the limited investigation performed in this study. We expect that deep multilayer convolution maps will be a promising alternative to bypass the higher dimensionality needed in shallow multilayer convolutions. However, such deep layering should also be accompanied by globally optimizing the different convolution maps so that they achieve a common objective. Ongoing efforts include the exploration of deep multilayer convolution maps and their connections to deep neural networks for building Markov models. Other directions being pursued include embedding nonlinearity within sGP-based methods and online update of SVD basis for POD-convolution models.

## 6. Acknowledgements

The work was performed with financial support from the Oklahoma State University (OSU) and computing support from the OSU High Performance Computing Center.